\documentclass{article}

\usepackage{cite}
\usepackage{graphicx}
\usepackage{caption}
\usepackage{amssymb}
\usepackage{array}
\usepackage{dcolumn}
\usepackage{bm}
\usepackage{textgreek}
\usepackage{amsmath}
\usepackage[letterpaper, total={7in, 9in}]{geometry}
\usepackage{authblk}
\usepackage{cleveref}
\crefformat{equation}{Eq.~(#2#1#3)}
\crefrangeformat{equation}{Eqs.~(#3#1#4--#5#2#6)}

\title{Multiplexing Guided Optical and Acoustic Waves for Efficient Acousto-Optic Devices}

\author[1,*]{Nathan Dostart}
\author[2]{Milo\v s Popovi\'c}

\affil[1]{Department of Electrical, Computer, and Energy Engineering, University of Colorado, Boulder, CO, 80309, USA}
\affil[2]{Department of Electrical and Computer Engineering, Boston University, Boston, MA, 02215, USA}

\affil[*]{Corresponding author: nathan.dostart@colorado.edu}

\date{\today}

\begin{document}

\maketitle

\begin{abstract}
Acousto-optic devices utilize the overlap of acoustic and optical fields to facilitate photon-phonon interactions. For tightly confined optical and acoustic fields, such as the sub-wavelength scales achievable in integrated devices, this interaction is enhanced. Broadband operation which fully benefits from this enhancement requires light and sound to co-propagate in the same cross-section, a geometry currently lacking in the field. We introduce the `acoustic-optical multiplexer', which enables this co-linear geometry, and demonstrate through simulations a proof-of-concept design. Using suspended silicon and silica beams, the multiplexer combines two optical modes and an acoustic mode into a single, co-guided output port with low insertion loss and reflection for both optics and acoustics. The first design in its class, the multiplexer enables integrated acousto-optic devices to achieve efficient photon-phonon interactions.
\end{abstract}

\section{Introduction}
\label{sec:intro}

Acousto-optic (AO) interactions, referred to in various fields as opto-mechanics, Brillouin or Raman scattering, or photon-phonon interactions \cite{van2016unifying}, provide a useful tool set for designing high performance optical devices. This is particularly true in integrated photonics where both optical and acoustic modes can be confined in sub-wavelength cross-sections. Recently, intense interest in bringing AO-based devices to integrated photonics have yielded isolators \cite{poulton2012design,kittlaus2018non,sohn2018time,ruesink2018optical}, amplifiers and lasers \cite{otterstrom2018silicon,gundavarapu2019sub,kittlaus2016large,kabakova2014chalcogenide,choudhary2017advanced,van2015net}, microwave photonic devices \cite{kittlaus2018rf,marpaung2015low,choudhary2017advanced}, modulators \cite{kittlaus2018non,sohn2018time,balram2016coherent,li2015nanophotonic,fan2016integrated}, and even optical phased arrays \cite{sarabalis2018optomechanical}.

Of principal importance in any AO device is maximizing the overlap of the optical and acoustic modes and thereby maximizing the AO interaction strength. Confining the acoustic mode to the same cross-section as the optical mode, and propagating along the same axis (co- or counter-propagating), allows \emph{all} of the acoustic and optical energy to be used for the AO interaction. The first demonstration of an AO tunable filter took advantage of such a co-propagating design \cite{harris1969acousto}, while it took 5 more years for demonstration of a non-collinear filter \cite{chang1974noncollinear}. Later demonstrations of AO filters in thin film LiNbO$_3$ \cite{kuhn1971optical}, implanted LiNbO$_3$ waveguides \cite{smith1990integrated}, and optical fiber \cite{kim1997all} among other platforms have also benefited from co-propagation of the optical and acoustic modes. While analogous devices have been demonstrated in integrated platforms where a transducer is used to drive the acoustic wave \cite{van2018electrical,sohn2018time,balram2016coherent,li2015nanophotonic,sarabalis2018optomechanical,fan2016integrated}, this co-propagating arrangement has not been demonstrated without utilizing an optical resonator or inducing excess optical loss due to the presence of the transducer.
The issue of efficiently coupling a transducer to a guided acoustic mode in the same cross-section with a guided optical mode is therefore an unsolved problem which, if addressed, would enable more efficient AO interactions in transducer-based designs.

Transducer-based approaches cannot directly drive a propagating acoustic wave in the same cross-section as an optical mode because the metal transducers will scatter the optical mode and induce loss.
The transducer-based designs demonstrated so far solve this issue by placing the transducer outside the optical waveguide and create the AO overlap in one of two manners: 1) launching an acoustic wave via a slab towards a waveguide   \cite{sohn2018time,balram2016coherent,li2015nanophotonic,sarabalis2018optomechanical} or 2) directly modulating the optical waveguide with standoff electrodes \cite{fan2016integrated,van2018electrical}. This first category can be further subdivided by whether the slab mode couples to a co-propagating, confined acoustic mode \cite{balram2016coherent,sarabalis2018optomechanical} or a `transverse' acoustic mode which is not confined to the waveguide and does not propagate significantly outside the transducer region \cite{sohn2018time,li2015nanophotonic}.
Slab-excitation of a co-propagating acoustic mode creates the desired geometry, but requires off-chip optical coupling \cite{balram2016coherent} or optical decay \cite{sarabalis2018optomechanical} in order to create strong AO overlap without the optical mode(s) scattering off the transducer.
Slab-excitation of a waveguide-transverse acoustic mode naturally prohibits coupling to confined, propagating acoustic mode since the confined wave by definition does not couple to radiation modes in the slab.
Direct modulation of the optical waveguide allows for near-optimal mode overlap and therefore strong AO interactions but requires large voltages as the electrodes must be sufficiently far from the waveguide and does not generate propagating acoustic modes, therefore requiring transducers as long as the desired interaction region \cite{fan2016integrated,van2018electrical}.
In order to enable this co-confined, co-propagating AO configuration we require a component which couples an external transducer to a confined, propagating acoustic mode in an optical waveguide without excess optical loss, reflection, or modal cross-talk.


\begin{figure*}[t]
	\centering
	\includegraphics[width=\textwidth]{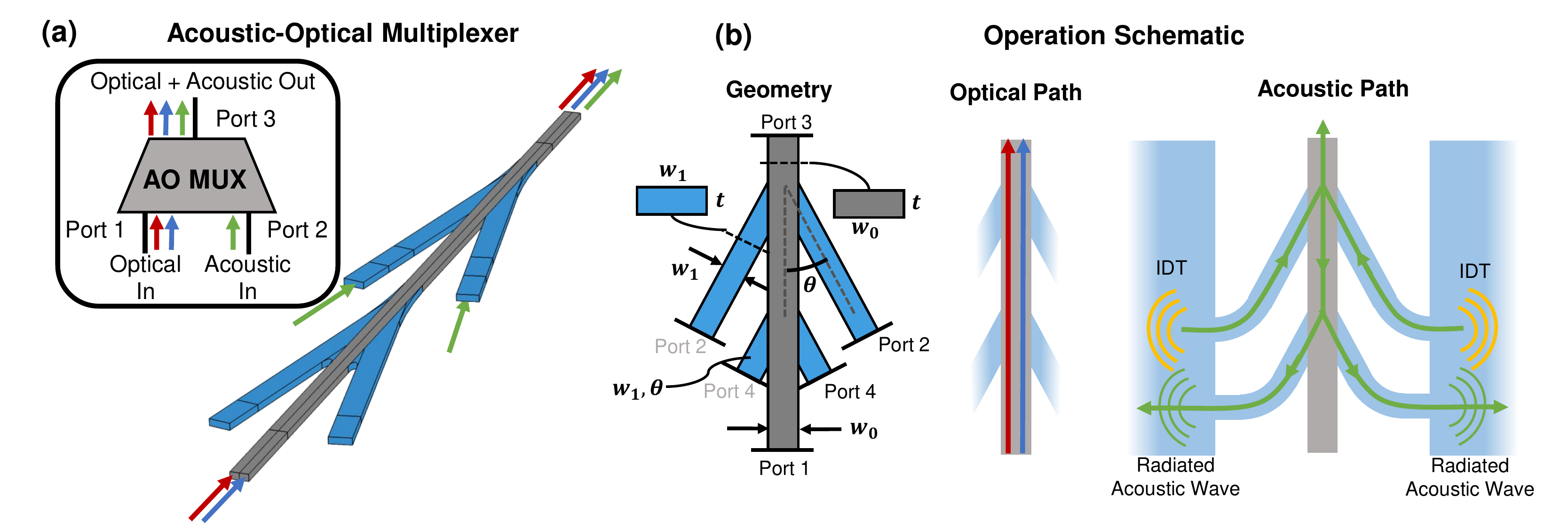}
	\caption{Overview of the acoustic-optical multiplexer. (a) Our example design multiplexes two optical modes (Port 1) in a suspended silicon waveguide with an acoustic wave incident symmetrically from suspended silica beams (Port 2) into a single output port, the suspended silicon beam (Port 3). (b) The geometry and port definitions of this architecture are shown along with schematic examples of the optical and acoustic paths. An additional pair of silica beams is used to redirect acoustic energy scattered in the backwards direction (Port 4).}
	\label{fig:overview}
\end{figure*}

In this paper we introduce the concept for such a device which combines optical and acoustic modes from separate spatial ports into a single spatial output port, an acoustic-optical multiplexer (AO mux). We demonstrate a simple implementation of an AO mux using only silicon (Si) and silica (SiO$_2$) so as to be compatible with the standard 220 nm SOI silicon photonics platform. Using COMSOL for acoustic simulations and the Lumerical finite-difference time-domain solver (FDTD) for optical simulations, we evaluate the device's performance to show that an AO mux is feasible without large acoustic or optical insertion loss, cross-talk, or reflections. This AO mux is the first design capable of coupling an external acoustic transducer to a non-resonant, guided acoustic mode co-propagating with an optical mode without disturbing the optical characteristics, and will enable more efficient transducer-based integrated AO devices.

\section{Device Overview}
\label{sec:design}

An AO mux requires, in the most basic sense, separate input ports for optical and acoustic modes and a output port which contains both optical and acoustic modes. A simple circuit schematic of an AO mux with ports is shown in the insert of Fig.~\ref{fig:overview}(a), where the two spatially separated input ports denoted as Ports 1 and 2 are multiplexed into a single output port, Port 3. Port 1 contains the optical mode, potentially including many waveguide and frequency modes, and Port 2 contains the acoustic mode(s). We emphasize that different spatial, rather than modal, ports are required precisely because there is no extant on-chip device capable of separating or combining acoustic modes. An AO mux, operating in reverse as an AO demux, provides exactly this functionality. The AO mux should ideally be both optically and acoustically transparent, i.e. should be without reflection, avoid inducing cross-talk between optical modes or acoustic modes, direct all optical and acoustic energy without loss into the output port, and be sufficiently short that AO interactions within the AO mux can be neglected.

To demonstrate the viability of such an AO mux, and provide a reference point for performance, we choose a simple design which couples a single acoustic mode from a suspended silica beam into a suspended silicon beam with two optical modes [Fig.~\ref{fig:overview}(a)]. The geometry of this design is shown in the left section Fig.~\ref{fig:overview}(b), while the optical and acoustic paths are shown in the center and right sections of Fig.~\ref{fig:overview}(b). We are particularly interested in designing a `CMOS-compatible' AO mux, motivated by demonstrations of acoustic transduction without piezoelectrics \cite{van2018electrical,weinstein2010resonant}, so we choose the materials and device layer thickness to be compatible with the standard 220 nm SOI integrated photonics platform: silica (blue) and silicon (gray) with thickness $t=220$ nm. Notably, silica is a `high index' material for acoustics but a low index material for optics, and the reverse is true for silicon. Any core-cladding structure which guides an acoustic mode will therefore anti-guide an optical mode, and vice versa. We choose silica beams to couple into/out of the acoustic mode of the silicon beam because it introduces minimal optical perturbation (low optical index) while also guiding the acoustic wave (high acoustic index). A suspended silicon beam is used to as the input optical port and output (muxed) port so as to confine the optical modes (in the high index silicon) while the air gap will confine the acoustic modes, a commonplace approach in integrated AO devices \cite{van2015net,van2018electrical,li2015nanophotonic,sohn2018time}. Notably, a silicon slot waveguide could co-confine both \cite{van2014analysis}. Addition of suspended silica beams could be relatively straightforward using directive etching techniques such as reactive ion etching or ion beam milling in combination with a sacrificial layer beneath the device layer.

An example application of an AO mux is for an isolator, where the AO interaction can be used to non-reciprocally convert between two optical modes \cite{poulton2012design,kittlaus2018non,sohn2018time,dostart2018energy}. We choose to design an AO mux centered at 1550 nm which facilitates this non-reciprocal mode conversion using a silicon beam of width $w_0$ where the TE0 and TE1 optical modes are coupled by the in-plane shearing acoustic mode. This corresponds to a silicon beam width of $w_0=570$ nm where the optical modes have identical group index of 4.11 and phase-matched acoustic wavelength $1.25$ {\textmu}m (frequency 3.12 GHz) \cite{dostart2018energy}. The silica beams have width $w_1=700$ nm and an angle $\theta=11^\circ$ relative to the silicon beam chosen to maximize acoustic directionality ($|S_{32}|^2/|S_{12}|^2$), bandwidth, and transmission ($|S_{32}|^2$).


Unfortunately, corners are strongly scattering for acoustics and cannot be avoided in our suspended geometry without adding a third, mutually low index material. In this design, where we have prioritized optical over acoustic transparency, these corners are the fundamental loss mechanism for the acoustic mode and excite undesired acoustic modes radiating into all ports. To ensure large acoustic directionality, i.e. no acoustic radiation into the optical input port, we added silica `siphon' beams which remove the acoustic radiation propagating into the optical input port (with an `internal' port denoted as Port 4). These siphons have the same width and angle relative to the silicon beam as the input silica beams to maximize the absorption of the undesired radiation through reciprocity. To further suppress excitation of undesired modes, we have arranged the AO mux to symmetrically excite (with two separate transducers) the desired transverse shear mode, and we have added fillets to the silica/silicon junctions to minimize the scattering due to edges.
    
    

    


\section{Simulation and Results}
\label{sec:results}

\begin{figure}[t]
	\centering
	\includegraphics[width=.6\textwidth]{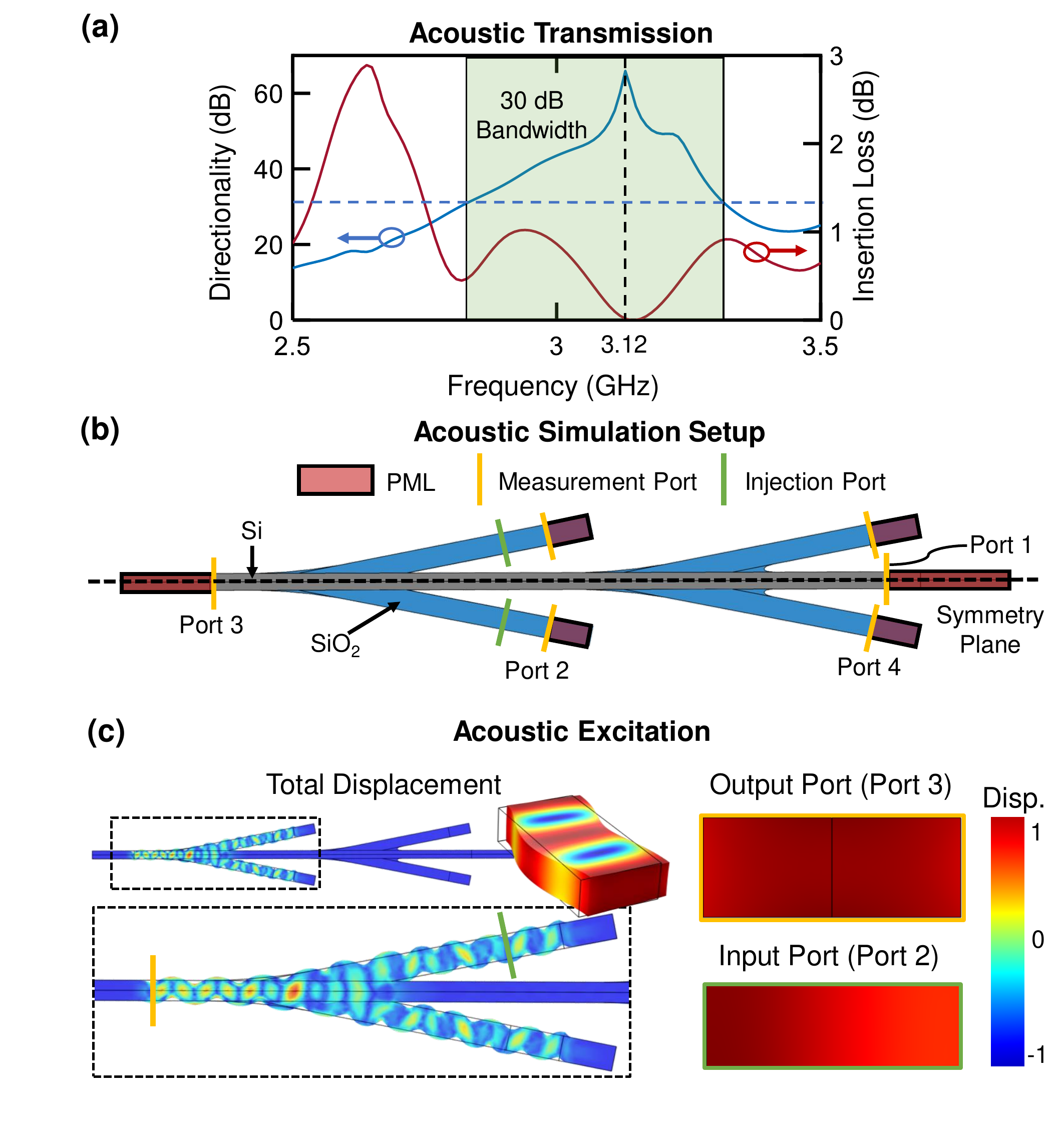}
	\caption{Acoustic performance of the AO mux. (a) Acoustic directivity and insertion loss for varying excitation frequency. (b) Acoustic simulation setup showing locations of ports, excitation planes, and PML locations. (c) Total displacement field at 3.12 GHz (left). Transverse displacement in the output port (gold) and excitation port (green) (right). Inset: 3D rendering of shear mode in the silicon beam. }
	\label{fig:acoustic_perf}
\end{figure}

\begin{figure}[t]
	\centering
	\includegraphics[width=.6\textwidth]{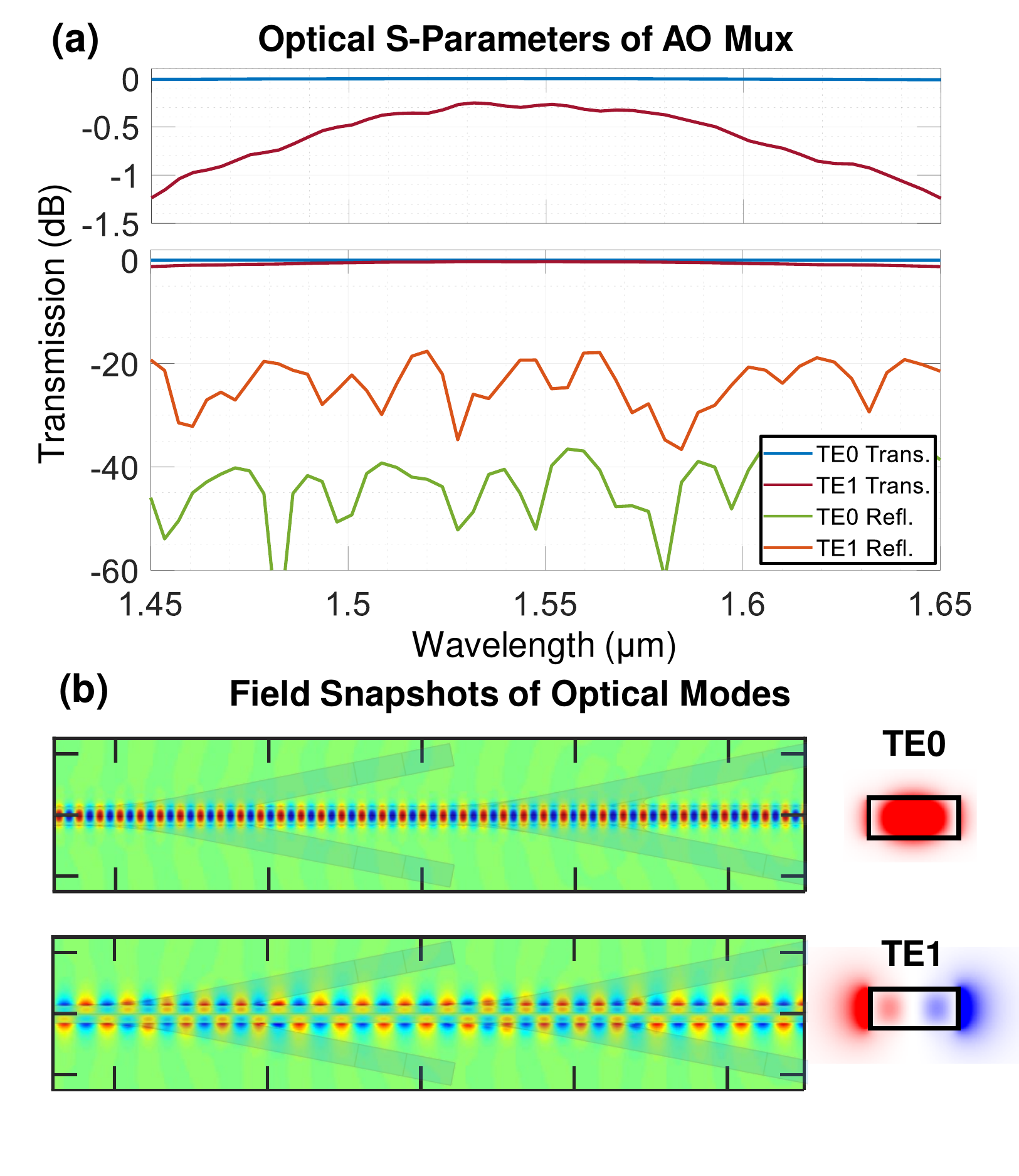}
	\caption{Optical performance of the AO mux. (a) Transmission and reflection of the TE0 and TE1 optical modes, with upper plot showing a zoomed view of the transmission. Mode crosstalk in both transmission and reflection is below $-70\,$dB. (b) Plots of the optical modes at 1550 nm propagating through the AO mux (left) and in cross-sectional view (right).}
	\label{fig:optical_perf}
\end{figure}

We demonstrate our AO mux design in simulation; in the acoustic domain, we use the COMSOL 3D Structural Mechanics module to implement the geometry and excite a single frequency acoustic wave with the Frequency Domain study module to measure the acoustic S-parameters. The results are shown in Fig.~\ref{fig:acoustic_perf}(a) for varying acoustic frequency. As can be seen, a `resonance' at the design frequency can be seen in the directionality arising from optimal matching between the incident mode in the silica beams and excited mode in the silicon beam. This matching ensures the backwards-going mode (into Port 1) is strongly suppressed and acoustic insertion loss is minimized. For the chosen geometry, the AO mux achieves a peak acoustic directionality of 60 dB and 0.05 dB insertion loss at the design frequency. It has a bandwidth of 500 MHz within which directionality is at least 30 dB and insertion loss varies between 0 and 1 dB. The simulation setup is shown in Fig.~\ref{fig:acoustic_perf}(b), where the measurement planes are denoted in gold and the symmetric excitation port is shown in green. A symmetry plane is applied to the center of the geometry, so only half of the geometry needs to be simulated. Perfectly matched layers (PMLs) are used as absorbing layers after each measurement port to avoid artificial reflections. To excite the acoustic wave, a forced displacement is applied to the excitation plane. The total power exiting from all measurement ports is used to calculate the total input power, normalize the power measured at each port, and calculate the acoustic S-parameters. Cross-sectional plots of the acoustic displacement, the shear field in the excitation port, and the shear field at the output port at center frequency are shown in Fig.~\ref{fig:acoustic_perf}(c), showing excitation of the desired mode into Port 3 with low insertion loss and high directionality.
   
For simulations in the optical domain, we import the geometry of the COMSOL simulation into Lumerical 3D FDTD. In two separate simulations, we excite pulses of TE0 and TE1 respectively and monitor Ports 1 and 3 as denoted in Fig.~\ref{fig:acoustic_perf}(b). Using a mode decomposition at both ports, the optical S-parameters are determined. The results are shown in Fig.~\ref{fig:optical_perf}(a), where the transmission and reflection coefficients are plotted across a 200 nm wavelength band centered at 1550 nm. The top plot is vertically scaled to show the insertion losses of both modes. Notably, TE0 has negligible insertion loss while TE1 has slightly larger insertion loss of varying between 1.2 dB at the edges of the optical bandwidth and 0.25 dB at 1550 nm. Both modes experience reflection $<-18$ dB and extremely low cross-talk of less than $-70$ dB, demonstrating that this AO mux design is effectively optically transparent. In Fig.~\ref{fig:optical_perf}(b) the modes are plotted in the device cross-section at 1550 nm, showing minimal perturbation by the silica beams.

We would like to emphasize that the presented design is not necessarily optimal, only a geometry which performs adequately and highlights the benefits and potential of an AO mux as well as some key considerations. For example, reflections in the TE1 optical mode and corresponding insertion loss could be considered the principal limitation of the current design. This could be improved by introducing the silica beams more adiabatically or widening the silicon beam in advance of the AO mux to better confine the TE1 mode. It should also be noted that the $<1$ dB acoustic insertion loss quoted here is within the mux, whereas additional insertion loss on the order of several dB is to be expected from simply coupling the transducer to the input acoustic port in the silica beam. Further improvements to this design would focus on simulation of the full system, including transducers, to minimize acoustic insertion loss as well as more complex beam shapes to minimize optical reflections.

Another potential improvement is to confine optical and acoustic waves in the same cross-section without suspension, already demonstrated in at least one geometry in an SOI platform using a `fin' waveguide \cite{sarabalis2017release}, a rib waveguide with a large height-to-width aspect ratio. Photonic-phononic crystals have, to our knowledge, required suspension for vertical acoustic confinement \cite{balram2016coherent}, but future improvements may allow for full confinement of both optics and acoustics without suspension using e.g. 1D vertical phononic crystals \cite{bahr2015theory}. Such non-suspended geometries could avoid the excess scattering due to corners by using a directional coupler configuration to evanescently couple an acoustic or optical mode from a second waveguide into the mux waveguide. In the presence of multiple optical modes, such as the case for our demonstrated design, an acoustic directional coupler would need to be used to avoid introducing optical mode cross-talk. For non-CMOS-compatible material sets, such as chalcogenides which can guide both optical and acoustic waves \cite{poulton2012design}, a cross-section where both optics and acoustics are guided can be easily designed and an AO mux should be well within reach.

In this paper we introduce the concept of an acoustic-optical multiplexer, an AO mux, which combines acoustic and optical waves from two separate spatial ports into a single co-guided cross-section for optimal overlap of acoustic and optical modes. The AO mux allows acoustic waves to be driven by a transducer and introduced to a co-guided waveguide without additional optical loss or use of a resonator. As a proof-of-concept device, we propose and demonstrate through simulations a simple AO mux design which multiplexes two optical modes in a silicon beam with an acoustic mode injected symmetrically from two silica beams. This basic geometry enables strong AO interactions without significant optical cross-talk or reflections while injecting the desired acoustic mode with a broad acoustic bandwidth of 500 MHz, $>30$ dB directionality, and at most 1 dB acoustic insertion loss. We expect that AO multiplexers will be a useful tool for designing high-efficiency transducer-based AO devices.

\noindent \textbf{\large Funding Information} National Science Foundation Graduate Research Fellowship Grant (1144083); Packard Fellowship for Science and Engineering (2012-38222).

\noindent \textbf{\large Acknowledgments} We thank Prof. Kelvin Wagner for discussions on previous demonstrations of AO devices.

\bibliographystyle{ieeetr}
\bibliography{biblio}

\end{document}